\documentclass[%
longbibliography,
reprint,
superscriptaddress,
nobibnotes,
 aps,
prstab,
]{revtex4-1}
\usepackage{ulem}
\usepackage[charter]{mathdesign}    % Math-design fonts
\usepackage{graphicx}% Include figure files
\usepackage{bm}% bold math
\usepackage[version=4]{mhchem}
\usepackage[svgnames]{xcolor}
\usepackage[
	colorlinks=True,linkcolor=DarkRed,citecolor=ForestGreen,urlcolor=MediumBlue,
	pdfstartview=FitH,bookmarks=False,pdfpagemode=UseNone
]{hyperref}
\renewcommand{\vec}{\mathbf}

\begin{document}

%\preprint{APS/123-QED}

\title{\boldmath Resonant Inelastic X-ray Scattering Study of Electron-Exciton Coupling in High-$T_c$ Cuprates}% Force line breaks with \\

\author{F.~Barantani}
 \affiliation{Department of Quantum Matter Physics, University of Geneva, 1211 Geneva, Switzerland}
 \affiliation{Institute of Physics, \'Ecole Polytechnique F\'ed\'erale de Lausanne, Lausanne, 1015, Switzerland}
\author{M.~K.~Tran}%
 \affiliation{Department of Quantum Matter Physics, University of Geneva, 1211 Geneva, Switzerland}%
\author{I.~Madan}
 \affiliation{Institute of Physics, \'Ecole Polytechnique F\'ed\'erale de Lausanne, Lausanne, 1015, Switzerland}
\author{I.~Kapon}%
 \affiliation{Department of Quantum Matter Physics, University of Geneva, 1211 Geneva, Switzerland}%
\author{N.~Bachar}
 \affiliation{Department of Quantum Matter Physics, University of Geneva, 1211 Geneva, Switzerland}%
  \affiliation{Department of Physics, Ariel University, Ariel, Israel}%
\author{T.~C.~Asmara}
\affiliation{Photon Science Division, Paul Scherrer Institut, 5232 Villigen PSI, Switzerland}%
\author{E.~Paris}
\affiliation{Photon Science Division, Paul Scherrer Institut, 5232 Villigen PSI, Switzerland}%
\author{Y.~Tseng}
\affiliation{Photon Science Division, Paul Scherrer Institut, 5232 Villigen PSI, Switzerland}%
\author{W.~Zhang}
\affiliation{Photon Science Division, Paul Scherrer Institut, 5232 Villigen PSI, Switzerland}%
\author{Y.~Hu}
\affiliation{Department of Applied Physics, Stanford University, CA 94305, USA}
\affiliation{Stanford Institute for Materials and Energy Sciences, SLAC,CA 94025, USA}
\author{E.~Giannini}
\affiliation{Department of Quantum Matter Physics, University of Geneva, 1211 Geneva, Switzerland}
\author{G.~Gu}
\affiliation{Brookhaven National Laboratory, Upton, NY 11973 5000, USA}
\author{T.~P.~Devereaux}
\affiliation{Stanford Institute for Materials and Energy Sciences, SLAC,CA 94025, USA}
\affiliation{Department of Materials Science and Engineering, Stanford University, Stanford, CA 94305, USA}
\affiliation{Geballe Laboratory for Advanced Materials, Stanford University, CA 94305, USA}
\author{C.~Berthod}
\affiliation{Department of Quantum Matter Physics, University of Geneva, 1211 Geneva, Switzerland}
\author{F.~Carbone}
\affiliation{Institute of Physics,  \'Ecole Polytechnique F\'ed\'erale de Lausanne, Lausanne, 1015, Switzerland}
\author{T.~Schmitt}
\affiliation{Photon Science Division, Paul Scherrer Institut, 5232 Villigen PSI, Switzerland}
\author{D.~van~der~Marel}
\affiliation{Department of Quantum Matter Physics, University of Geneva, 1211 Geneva, Switzerland}

\date{\today}

\begin{abstract}
Explaining the mechanism of superconductivity in the high-$T_c$ cuprates requires an understanding of what causes electrons to form Cooper pairs.
Pairing can be mediated by phonons, the screened Coulomb force, spin or charge fluctuations, excitons, or by a combination of these.
An excitonic pairing mechanism has been postulated, but experimental evidence for coupling between conduction electrons and excitons in the cuprates is sporadic.
Here we use resonant inelastic x-ray scattering (RIXS) to monitor the temperature dependence of the $\underline{d}d$ exciton spectrum of Bi$_2$Sr$_2$CaCu$_2$O$_{8-x}$ (Bi-2212) crystals with different charge carrier concentrations.
We observe a significant change of the $\underline{d}d$ exciton spectra when the materials pass from the normal state into the superconductor state.
Our observations show that the $\underline{d}d$ excitons start to shift up (down) in the overdoped (underdoped) sample when the material enters the superconducting phase. 
We attribute the superconductivity-induced effect and its sign-reversal from underdoped to overdoped to the exchange coupling of the site of the $\underline{d}d$ exciton to the surrounding copper spins.
\end{abstract}

\maketitle

\noindent

\section*{Introduction}

Ever since Bednorz and Muller's~\cite{bednorz1986} discovery of high-$T_c$ superconductivity in cuprates the question as to what mediates superconductivity in these materials has continued to occupy the scientific community. The strong electron correlations of the Cu $3d$ states form a major theoretical challenge~\cite{anderson1987}. In particular, in the insulating parent compounds the on-site part of the Coulomb repulsion $U$ splits the Cu~$3d$ band into a fully occupied lower Hubbard band (LHB) and an empty upper Hubbard band (UHB). 
The LHB and UHB are $\sim 8$~eV apart and the O~$2p$ band, which is fully occupied, falls \textit{inside} this gap with the top of the O~$2p$ band about $2$~eV below the UHB. The separation between the top of the O~$2p$ band and the UHB, the so-called charge transfer gap $\Delta_{CT}$~\cite{zaanen1985,ghijsen1988}, constitutes the lower bound of the electron-hole continuum.

\begin{figure*}[!t]
\centering
\includegraphics[width=0.8\textwidth]{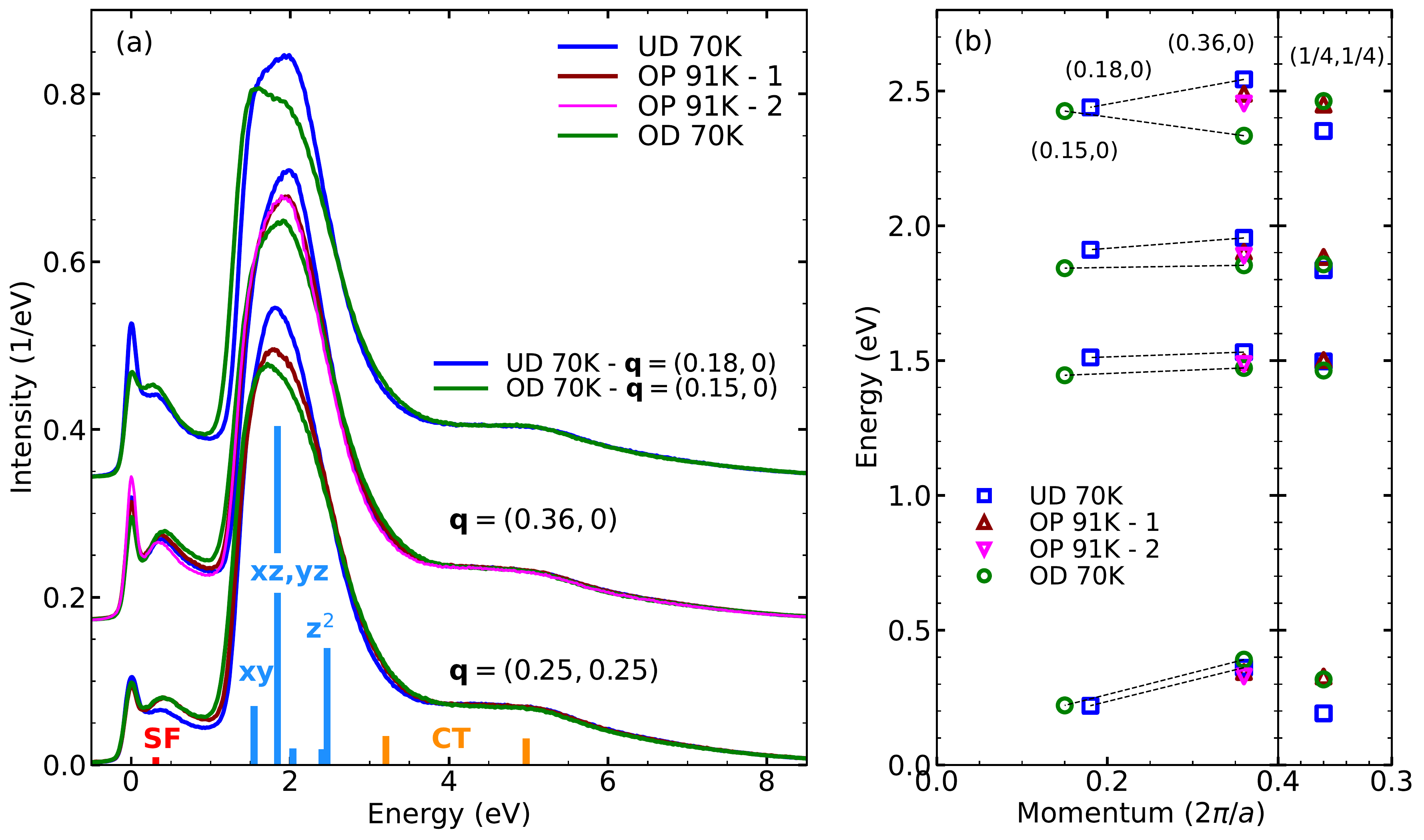}
\caption{\small (a) Experimental RIXS spectra for the three different dopings and for different momentum transfers. To avoid clutter the spectra have been given vertical offsets grouped according to the momentum values. Vertical bars correspond to spin flip (red), $\underline{d}d$ excitons (blue) and charge transfer excitations (orange) obtained from a cluster calculation (see Appendix~\ref{app:multipletRIXS}). (b) Energy momentum dispersion of the fitted peak positions. Error bars are smaller than the symbol size and therefore not indicated. Momentum values are indicated on top. 
}
\label{fig:RIXS1}
\end{figure*}
Doping either electrons or holes induces an insulator-to-metal transition. 
It has turned out to be difficult to obtain a simple analytical framework describing simultaneously the localized charges of the correlated insulator and the mobile charges responsible for the metallic conductivity. 
Even the Hubbard model, describing a single band of charge carriers with on-site Coulomb repulsion $U$, can not be solved analytically in 2 or 3 dimensions, and the properties and predictions of this model are based on the numerical solution of small clusters via density matrix renormalization group and quantum Monte Carlo, or approximation schemes such as dynamical mean field theory ~\cite{arovas2021}. 
The present status of the Hubbard model is that in 2 dimensions the superconducting state is unstable with respect to charge stripes, and that stable superconductivity requires the suppression of the charge stripes state by introducing next-nearest neighbor hopping, longer range hoppings, dynamical lattice effects and other orbital contributions~\cite{jiang2019}.

Theoretical studies of superconductivity in cuprates have mostly concentrated on the aforementioned Hubbard model and the related $t$--$J$ model for which the fundamental interaction is non-retarded~\cite{anderson2007}.
When interacting holes form Cooper pairs, the effective-mass change resulting from the pair formation may stabilize the superconducting state~\cite{hirsch2002}. On a microscopic scale a pairing-induced modification of the bare density response function~\cite{leggett1999}, magnon mediated-interactions~\cite{eschrig2006,scalapino2012}, plasmon-mediated interactions~\cite{bill2003}, pairing mediated by the ionic electron polarizabilities~\cite{sawatzky2009,yacoby2016}, and exciton-mediated pairing~\cite{weber1988,zaanen1993,feiner1992,buda1994,holcomb1994,bucci1995,little2007,mazov2012} derive from the Coulomb interaction between the electrons (for a short review see Ref.~\cite{oles2019}), which does not involve lattice degrees of freedom in contrast to phonon-mediated pairing~\cite{maksimov2010}. 
The concept of exciton-mediated pairing has been pioneered in the context of organic materials~\cite{little1964}, and sandwich structures~\cite{ginzburg1970b,allender1973,mazov2012}. It has also been considered for the cuprates~\cite{weber1988,zaanen1993,feiner1992,buda1994,holcomb1994,bucci1995,little2007,mazov2012}, but experimental information has remained scarce. 
Excitons are bound states of an electron and a hole due to the attraction resulting from their opposite charge.
The strongest binding occurs when the electron and the hole occupy the same site, namely, when an electron from one of the occupied $d_{xy}$, $d_{yz}$, $d_{zx}$, or $d_{z^2}$ states is excited into the partially filled $d_{x^2-y^2}$ state. Since this does not change the site occupancy there is no energy cost of order $U$ and the only contribution comes from the level splitting due to the crystal field~\cite{zaanen1993,wang2011}. 
Such $\underline{d}d$ excitons are the most prominent features of resonant inelastic x-ray scattering (RIXS) spectra at the Cu $L$-edge~\cite{kuiper1998} and from numerous experimental~{\cite{ghiringhelli2004,ghiringhelli2007,morettisala2011,dean2013,dean2014}} and theoretical~\cite{wang2011} studies it has become clear that they are robust for all experimentally achievable carrier concentrations.

It is known that phonons coupled to conduction electrons change energy and lifetime when entering the superconducting phase~\cite{axe1973,shapiro1975,altendorf1993}. This behavior is attributed  to a change of the self-energy of the phonons below $T_c$ due to the fact that they are coupled to the conduction electrons~\cite{zeyher1990}. 
On the other hand, when a $\underline{d}d$ exciton is created, there is an orbital flip from the $3d_{x^2-y^2}$ to one of the other Cu $3d$ levels. Thus, the $\underline{d}d$ excitons are firmly tied to the carriers at the Fermi level. 
Here we report the observation of the renormalization of $\underline{d}d$ excitons with energies in the range of $1$--$3$~eV. Using RIXS we observe the impact of the superconducting order on the $\underline{d}d$ excitons in \ce{Bi2Sr2CaCu2O_{8-x}}.
\begin{figure*}
\centering
\includegraphics[width=0.9\textwidth]{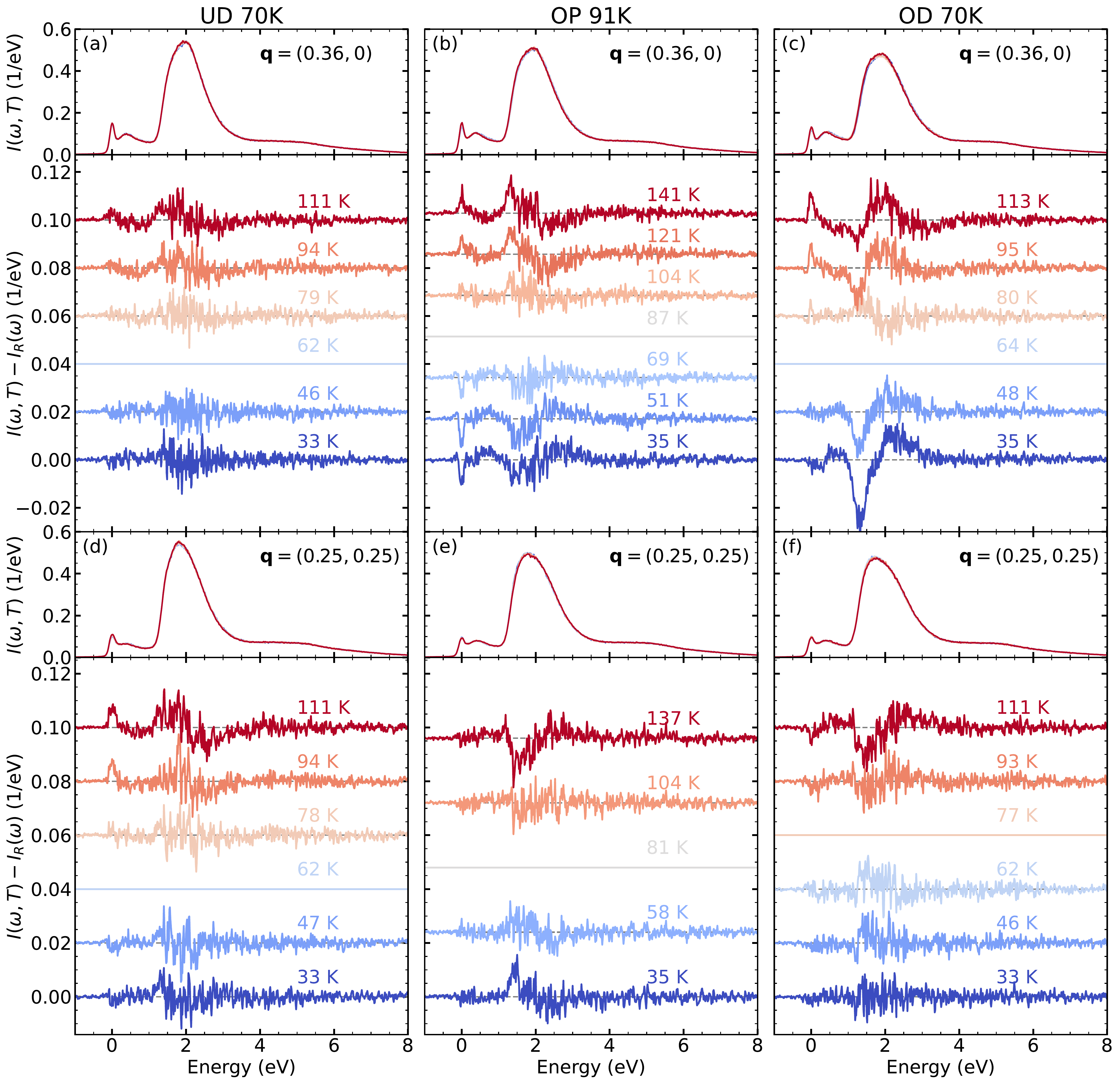}
\caption{\small 
Top panels: Normalized RIXS spectra: the spectra for different temperatures overlap except near the maximum of the $\underline{d}d$ peak. 
Bottom panels: Normalized RIXS spectra for the same temperatures as the top panels from which the spectra at $T_c$ have been subtracted.  
(a,c) Underdoped Bi-2212 ($T_c=70$~K). 
(b,e) Optimally doped Bi-2212 ($T_c=91$~K).
(c,f) Overdoped Bi-2212  ($T_c=70$~K).
(a,b,c) Momentum value $(0.36,0)$. 
(d,e,f) Momentum value $(0.25,0.25)$.
}
\label{fig:spectra}
\end{figure*}

\section*{Experimental results}

\noindent
We measured RIXS spectra of crystals from the \ce{Bi_2Sr_2CaCu_2O_{8-x}} bi-layer cuprate family across the superconducting dome: underdoped with carrier concentration $p=0.11$ and $T_c=70$~K, optimally doped ($p=0.16$, $T_c=91$~K), and overdoped ($p=0.21$, $T_c=70$~K).
The measurements, performed at the ADRESS beamline of the Swiss Light Source at the Paul Scherrer Institut (PSI)~\cite{strocov2010,ghiringhelli2006}, were carried out using $\sigma$-polarization of the incoming x-rays with energy resonant at the Cu $L_3$ edge. 
We express the momentum vectors on the basis of the pseudo-tetragonal structure with planar lattice parameter $a=3.83$~\AA{} along the CuO bonds and $c=30.9$~\AA{} perpendicular to the planes. 
Each crystal was aligned on the sample holder using Laue diffraction, and data were collected for selected values of the momentum transfer $\vec{q}=(\mu,\nu,j)$ in units of the reciprocal lattice, where $j$ is an integer number. Keeping in mind the periodicity in reciprocal space, we will use the shorthand notation $(\mu,\nu)$.
\begin{figure*}[!t]
\centering
\includegraphics[width=0.9\textwidth]{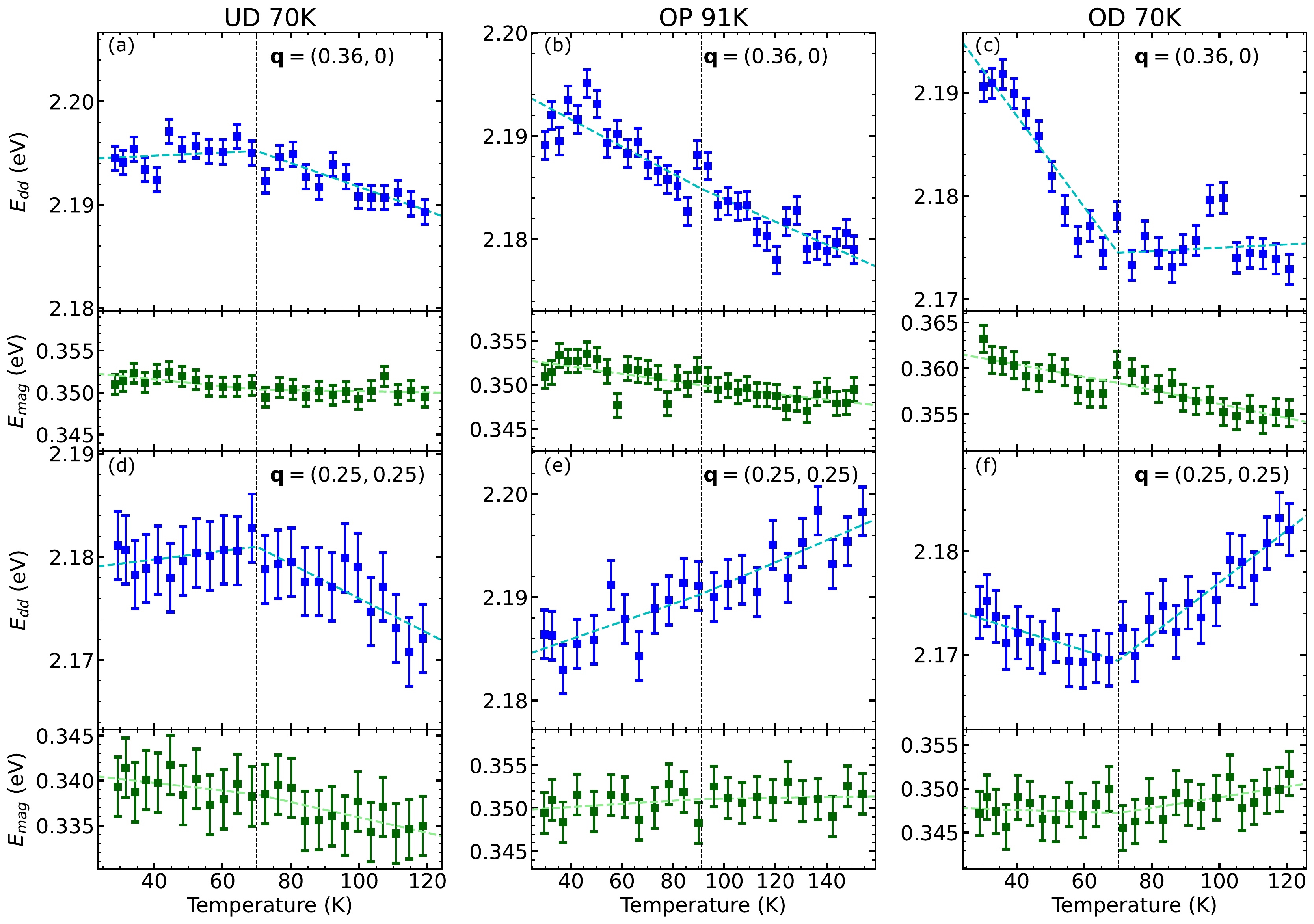}
\caption{\small Temperature dependence of the first moment of the $\underline{d}d$ exciton peak and the magnon peak, as extracted from the experimental spectra (considering respectively the energy intervals $1$--$4$~eV and $0.1$--$0.6$~eV). The dashed lines are fits to a curve parametrized as Eq.~(\ref{eq:fit}). The error bars represent the uncertainty on the elastic peak position computed as explained in Appendix~\ref{section:data_acquisition}. The vertical dotted lines indicate $T_c$.
Top panels: momentum value $(0.36,0)$. 
Bottom panels: momentum value $(0.25,0.25)$.
(a,d) Underdoped Bi-2212 ($T_c= 70$~K). (b,e) Optimally doped Bi-2212 ($T_c= 91$~K). (c,f) Overdoped Bi-2212 ($T_c= 70$~K).}
\label{fig:first_moment}
\end{figure*}
\begin{figure*}[!t]
\centering
\includegraphics[width=0.7\textwidth]{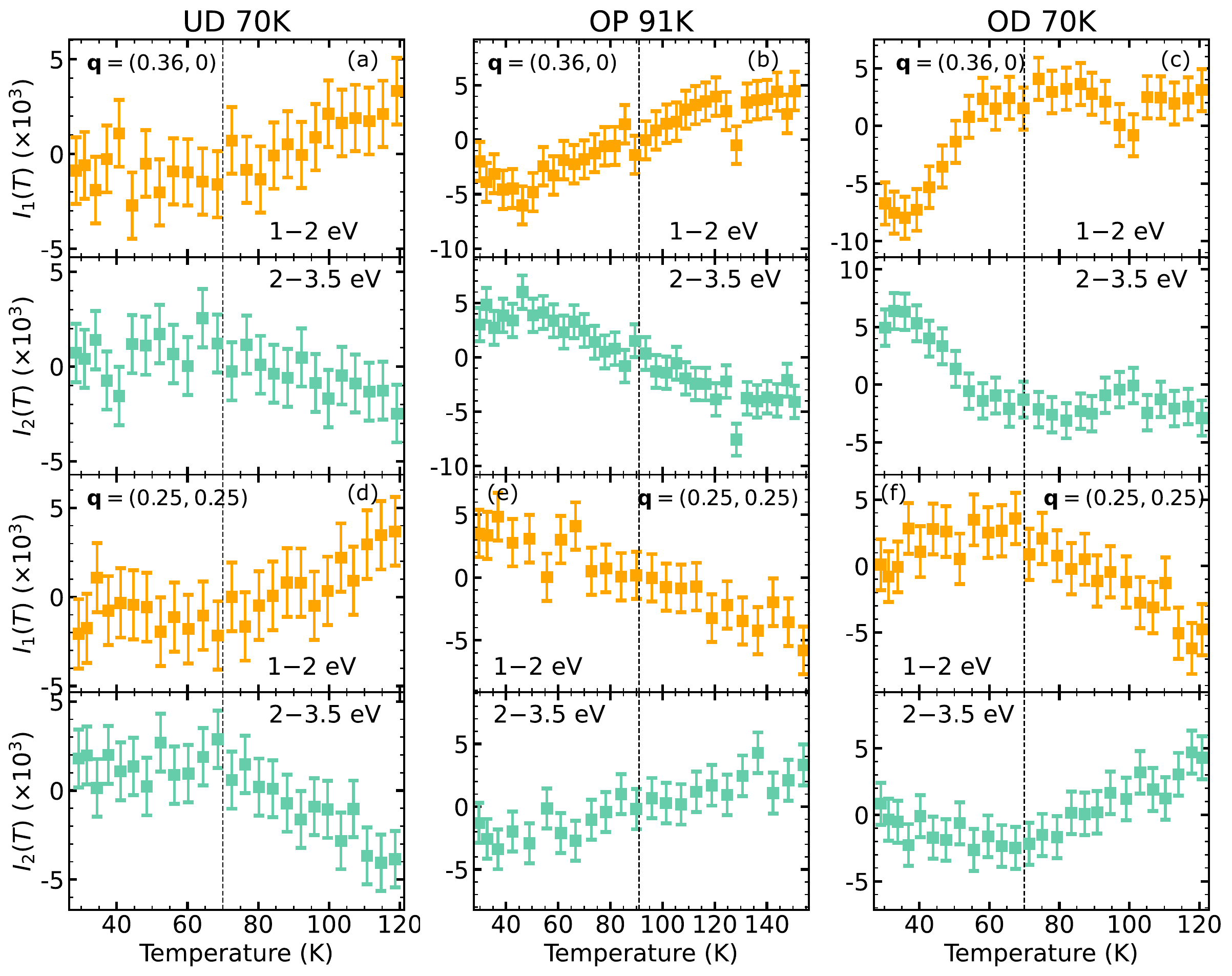}
\caption{\small Temperature dependence of the integrated RIXS intensity $I_j(T)$ where the index $j=1,2$ refers to the two energy ranges $1$--$2$~eV, and $2$--$3.5$~eV respectively. The dashed lines indicate $T_c$. 
Top panels: momentum value $(0.36,0)$. 
Bottom panels: momentum value $(0.25,0.25)$.
(a,d) Underdoped Bi-2212 ($T_c= 70$~K). (b,e) Optimally doped Bi-2212 ($T_c= 91$~K). (c,f) Overdoped Bi-2212 ($T_c= 70$~K).
}
\label{fig:tot_counts}
\end{figure*}
The samples were cleaved at $20$~K in ultra high vacuum (UHV) directly on the manipulator for RIXS data acquisition.

In Fig.~\ref{fig:RIXS1}(a) we compare RIXS spectra of Bi-2212 for different dopings and momentum values. Since as we will see below, the temperature dependence is too weak to be noticeable on this scale, for each given doping $p$ and momentum value $\vec{q}$ the spectra $I_{R}(p,\vec{q},\omega)$ were averaged in the temperature range of approximately $50$~K above and below $T_c$. 
The vertical bars are theoretical predictions of the excitation energies following from a cluster calculation (see Appendix~\ref{app:multipletRIXS}).
We see that the central peak at $1.9$~eV corresponds to the $\underline{d}_{xz}d_{x^2-y^2}$ and $\underline{d}_{yz}d_{x^2-y^2}$ excitons, the shoulder at $1.5$~eV to $\underline{d}_{xy}d_{x^2-y^2}$ and the shoulder at $2.4$~eV to $\underline{d}_{z^2}d_{x^2-y^2}$. 
We have fitted the spectra to a linear superposition of Voigt profiles (see Appendix~\ref{section:data_acquisition}). 
The $\underline{d}d$ manifold is described by the three aforementioned peaks, with Gaussian profiles having a FWHM of around (0.5, 0.9, 0.4)~eV respectively for (1.5, 1.9, 2.4)~eV. This line shape has been shown by Lee {\it et al.} to originate in the Franck-Condon principle, resulting in phonon overtones of the electronic mode described by a Gaussian envelope~\cite{lee2014}.
We observe that, when the doping increases, the $\underline{d}d$ excitons shift to lower energy. In particular, from the peak positions extracted by fitting, we can see that for $\vec{q}=(0.36,0)$, where the effect is strongest, all excitons shift to lower energy.

The momentum dispersion for the three different dopings is shown in Fig.~\ref{fig:RIXS1}(b). We see that the peak positions have relatively small energy-momentum dispersion, which confirms the observations of Moretti \textit{et al.}~\cite{morettisala2011}.
This finding underlines the strongly localized character of these excitations. The peak energy below 0.5 eV has strong momentum dependence, which is expected for a magnon. 

In the top panels of Fig.~\ref{fig:spectra} we present the RIXS data for different dopings, momentum transfers and for different temperatures. The data in this figure are averages spanning intervals of $15$~K (UD 70K and OD 70K) and  $20$~K (OP 91K) in order to minimize the statistical noise.
In order to highlight the spectral temperature dependence, we show in the bottom panels the spectra from which the spectrum taken at $T_c$ is subtracted.
The temperature dependence in the paramagnon region between $0.2$~eV and $1$~eV is very weak, which is consistent with the observations of Peng \textit{et al.}~\cite{peng2015} for optimally doped Bi$_{1.5}$Pb$_{0.55}$Sr$_{1.6}$La$_{0.4}$CuO$_{6+\delta}$ at $50$~K and $200$~K. 
Also the charge transfer region between 3.5 and 8~eV has negligible temperature dependence. 
The most significant temperature dependence occurs for the underdoped and overdoped samples in the part of the spectrum corresponding to the $\underline{d}d$ excitons.
In particular the difference spectra show that spectral weight is redistributed between different parts of the $\underline{d}d$ manifold as a function of temperature.  We calculated for each spectrum the average energy
\begin{equation}
E =
\frac{\int_{\omega_1}^{\omega_2}I(\omega) \omega d\omega }
{\int_{\omega_1}^{\omega_2}I(\omega) d\omega  }.
\label{eq:E}
\end{equation}
In Fig.~\ref{fig:first_moment} the temperature dependence of the magnon and exciton energies are displayed using Eq.~(\ref{eq:E}). For the magnon we use the integration limits $\omega_1=0.1$~eV and $\omega_2=0.6$~eV,  for the exciton $\omega_1=1$~eV and $\omega_2=4$~eV. 
In~Appendix~\ref{section:data_acquisition} we demonstrate that the temperature evolution is essentially a uniform shift of  $\underline{d}_{xy}d$ and $\underline{d}_{xz,yz}d$, with a significantly smaller shift of $\underline{d}_{z^2}d$. 
Since the latter component does not contribute very strongly to the overall spectral weight of the $\underline{d}d$ manifold, the average $\underline{d}d$  exciton energy  provides a meaningful representation of the thermal evolution of the spectrum.  
We see that this energy shifts as function of temperature for different dopings. The temperature dependence shows an upward change of slope at $T_c$ for the overdoped case [in Fig.~\ref{fig:first_moment}(c) and (f)], and a downward change with smaller magnitude in the underdoped sample [panels (a) and (d)]. In contrast the magnon peak shows a negligible, if not absent, superconductivity induced effect.
It is of interest to make a connection to the doping dependence shown in Fig.~\ref{fig:RIXS1}, where we noticed that for $\vec{q}=(0.36,0)$ the excitons soften with increasing doping. These trends are consistent with the evolution of either antiferromagnetic or singlet correlations as a function of doping and temperature. That said, for $\vec{q}=(0.25,0.25)$ both doping and temperature dependence appear to follow a more complex pattern. 

These temperature dependencies are also clearly revealed by the intensities of the left-hand side (1--2~eV) and right-hand side (2--3.5~eV) of the $\underline{d}d$ manifold, confirming the change of slope near $T_c$ for the underdoped and overdoped samples (see Fig.~\ref{fig:tot_counts}).
To quantify these slope changes we have fitted the temperature dependence of the average energies of the magnon and the exciton manifold to
\begin{align}
&T\ge T_c:  E(T)=E_c+E_{n}\left(1-T/T_c\right)
\nonumber \\
&T\le T_c:  E(T)=E_c+\left(E_{n}+E_{sc}\right)\left(1-T/T_c\right)
\label{eq:fit}
\end{align}
by adjusting the normal state parameters $E_c$, $E_n$, and the superconductivity induced change of energy $E_{sc}$.  The dashed curves in Fig.~\ref{fig:first_moment} are least-squares fits of this expression to the experimental data points. 
In Fig.~\ref{fig:tot_counts} we present the temperature dependence of the integrated intensity of the spectra shown in Fig.~\ref{fig:spectra}
\begin{equation}
I_{j}(p,\vec{q},T) = \int_{\omega_j}^{\omega_{j}^{\prime}}\left[I(p,\vec{q},\omega,T)-I_{R}(p,\vec{q},\omega)\right] d\omega
\end{equation}
in two regions defined by $(\omega_j,\omega_{j}^{\prime})$: $1$--$2$~eV (region~1), and $2$--$3.5$~eV (region~2).
The split at 2~eV is motivated by the fact that for all panels in Fig.~\ref{fig:spectra} the intensity on either side has opposite temperature dependence. It thus provides optimal signal-to-noise ratio of the temperature dependence in the $\underline{d}d$ exciton region.
Comparing the intensity in regions 1 and 2, we see that, when entering the superconducting phase, for the overdoped case spectral weight is transferred from the low energy to the high energy side of the $\underline{d}d$ exciton spectrum and vice versa for the underdoped case. 
As in Fig.~\ref{fig:first_moment}, we observe a change of slope in the $\underline{d}d$ temperature dependence around $T_c$, except for the optimally doped sample.
As can be seen in Fig.~\ref{fig:first_moment}, for the OD 70K the contribution of the superconducting state is to shift the $\underline{d}d$ exciton peak to higher energies. This conclusion is consistent with the superconductivity-induced effect on the integrated intensity $I_{1,2}(T)$ which indicates a transfer of the spectral weight towards higher energies.

\begin{figure}[]
\centering
\includegraphics[width=0.95\columnwidth]{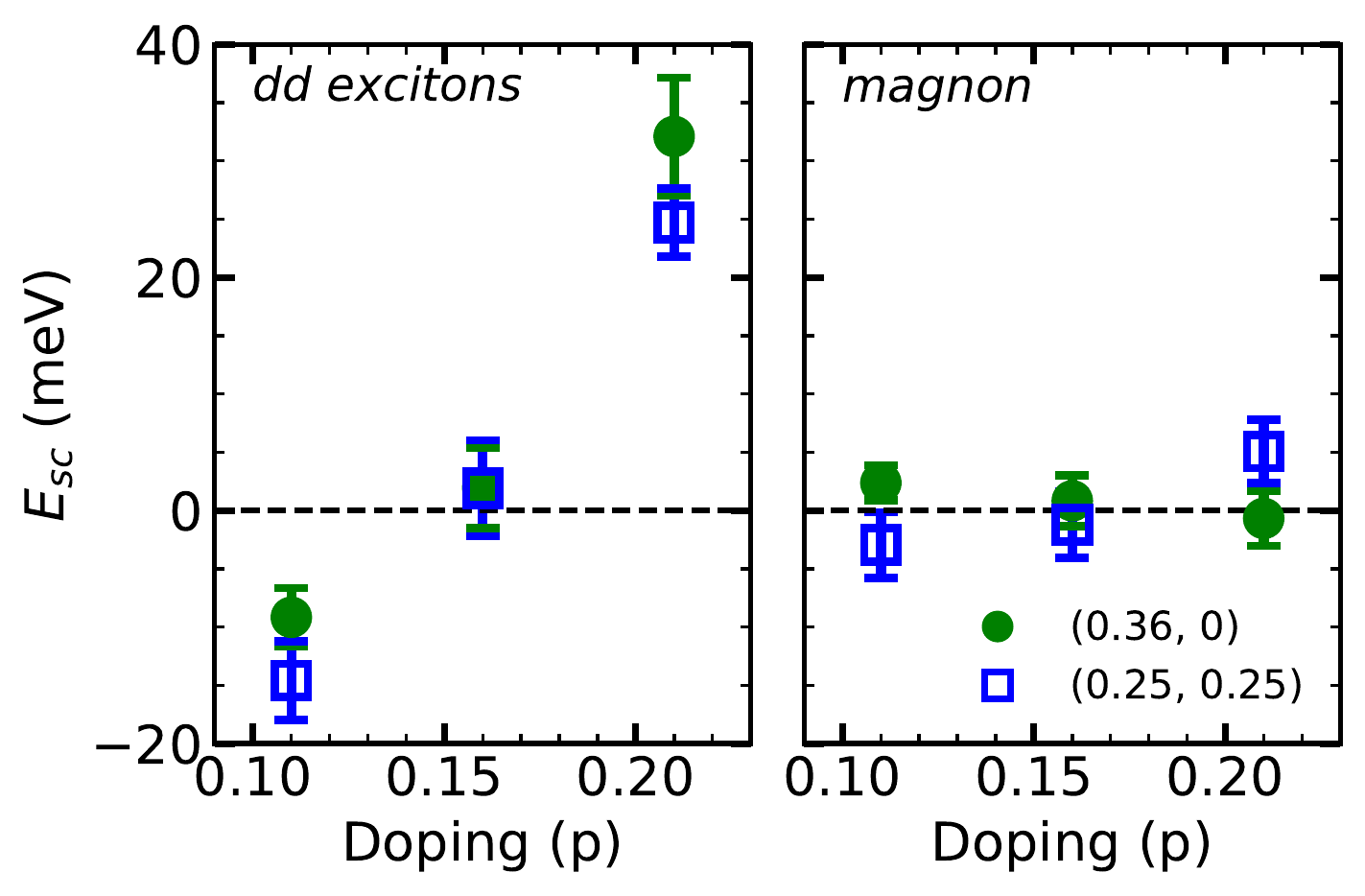}
\caption{\small Doping dependence of the superconductivity-induced change of the average $\underline{d}d$ exciton and magnon energies. The momentum values are $(0.36,0)$ (green circles) and $(0.25,0.25)$  (blue squares). The error bars are extracted from the fit procedure.}
\label{fig:coeff_SC}
\end{figure}

In Fig.~\ref{fig:coeff_SC} the superconductivity-induced energy changes $E_{sc}$ of the $\underline{d}d$ excitons and of the magnon are compared for all samples and momentum directions. 
At the overdoped side the effect of the superconducting order is to shift the excitons from lower to higher energy, both along the nodal and antinodal directions.  We observe the opposite trend for the underdoped sample, again for the nodal as well as the antinodal direction in momentum space. For the optimally doped sample the superconductivity-induced change is small along either direction. For the magnon, the influence of the superconducting phase transition, if it exists, falls within the experimental noise.

\section*{Discussion}
\begin{figure*}[]
\centering
\includegraphics[width=2\columnwidth]{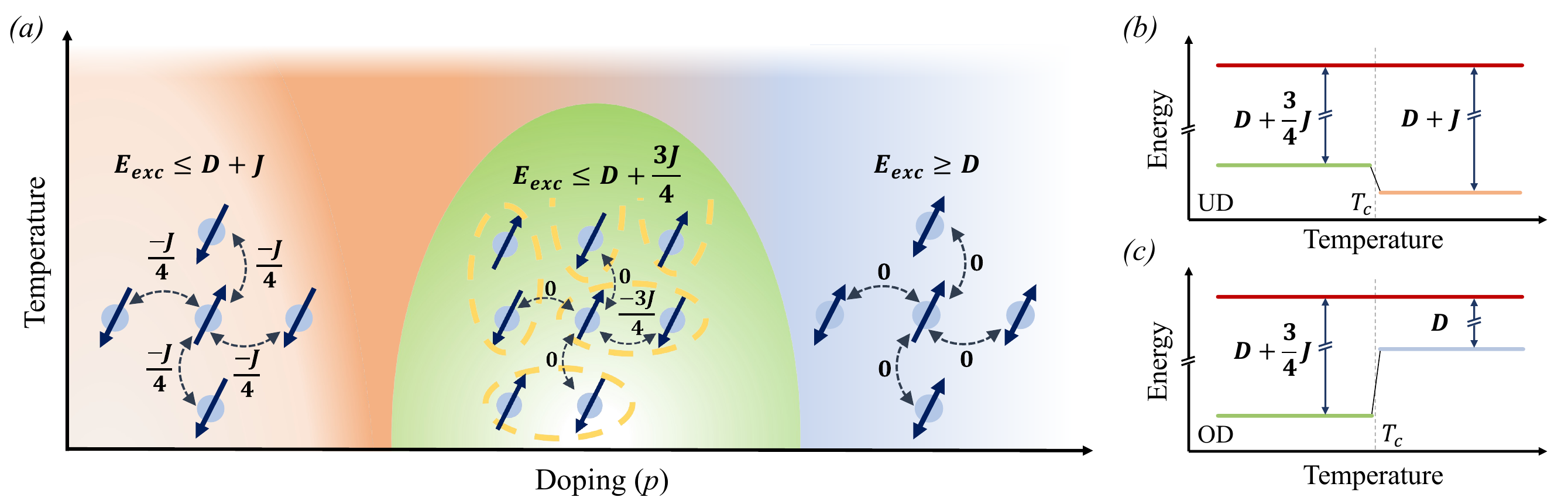}
\caption{\small 
(a) Phase diagram of the cuprates indicating the antiferromagnetic phase (beige), the normal phase (orange for underdoped and light blue for overdoped) and the superconducting phase (green) with the corresponding short-range spin correlations and $\underline{d}d$ exciton energies.
(b,c)  Sketch of the temperature dependence of the $\underline{d}d$ exciton energy in the underdoped and overdoped cases.
}
\label{fig:JJ}
\end{figure*}
The aforementioned electron-phonon coupling resulting in the Gaussian line shape~\cite{lee2014} can in principle lead to a finite temperature dependence of the $\underline{d}d$ excitons, if the coupled vibrational modes are sufficiently anharmonic. Such an effect has for example been observed in the vibronic spectra of BEH-PPV conjugated polymer films~\cite{oliveira2003}, where the excitons have an Arrhenius type shift to higher energy with increasing temperature. The observed temperature dependence in the present case of the $\underline{d}d$ excitons may, at least in part, be due to the electron-phonon coupling. A natural question is whether the observed change of slope at the superconducting phase transition would also be a consequence of the electron-phonon coupling. To have such a strong effect requires that the phonon anharmonicity is strongly influenced by the superconducting phase transition.
Refs.~\cite{ahn2021} and ~\cite{chen2021} discuss superconductivity induced change of orbital character and its impact on the optical spectra. With some modifications, similar notions come into play into the RIXS $\underline{d}d$ exciton spectra if, in conjunction with the SC phase transition, there is a change of orbital character of the bands. The influence on the excitonic spectrum is however very indirect and quantitatively much too small compared to our observations.
Little and Holcomb~\cite{holcomb1994,little2007} analyzed the optical conductivity in the superconducting and normal states and arrived at the conclusion that the charge carriers are coupled to the $\underline{d}d$ excitons.  
Vice versa one may expect a manifestation in the exciton spectrum of a mixing with the electron-hole continuum, causing a renormalization of the exciton energy $\omega_{\underline{d}d}$ by a self-energy $g_{\vec{q}}^2\chi(\vec{q},\omega)$. Here $\chi(\vec{q},\omega)$ is the charge susceptibility and $g_{\vec{q}}$ is a momentum dependent coupling constant. 
Husain {\it et al.}~\cite{husain2019} reported a strong temperature dependence of the imaginary part of the charge susceptibility at $\vec{q}=(0.17, 0.17)$, with a change of sign from underdoped to overdoped. In the weak coupling approach this phenomenology translates into a temperature dependent exciton energy having a strong doping dependence of the temperature coefficient. 
When the material passes from normal to superconducting, the susceptibility changes, causing both an energy shift and a change of the lifetime of the exciton, similar to what happens in the case of electron-phonon coupling~\cite{axe1973,shapiro1975,altendorf1993,zeyher1990}. The susceptibility function satisfies $V_{\vec{q}}\chi(\vec{q},\omega)=1/\epsilon(\vec{q},\omega)-1$, where $V_{\vec{q}}$  is the Coulomb potential and $\epsilon(\vec{q},\omega)$ the dielectric function. For $q\approx 0$ the dielectric function  $\epsilon(\vec{q},\omega)$  can be measured with great precision using optical techniques. Experimentally it was found that the electron-hole susceptibility of the cuprates near $q=0$ is significantly impacted by the superconducting order, even for energies higher than 1~eV~\cite{levallois2016}. 
By considering the renormalization due to the BCS charge susceptibility,  the temperature behavior calculated for the exciton is qualitatively similar to the experimental observations at the overdoped side (Appendix \ref{section:BCS}). Quantitatively this weak-coupling model falls short in describing the experimental data: even with the electron-exciton coupling constant set to $g\sim 2$~eV, the calculated shift is an order of magnitude smaller than in the experiment. 

An explanation may be found in the fact that the superconducting carriers move in the $3d_{x^2-y^2}$ bands, while a $\underline{d}d$ exciton is effectively a localized orbital flip of a $3d_{x^2-y^2}$ hole to a different $3d$ orbital. The circumstance that the electronic conductivity and the excitons involve the same electrons calls for a more radical approach than offered by the weak coupling scheme sketched above. In fact, if the material is superconducting the orbital flip of one of the $3d_{x^2-y^2}$ holes necessarily breaks up a Cooper's pair. Consequently when passing from the normal state to the superconducting state the exciton energy increases by the pair breaking energy. The observation of a weaker effect of the opposite sign for the underdoped sample appears to be, as we will argue below, a manifestation of the physics underlying the pseudogap in the cuprates. 
A deeper understanding is obtained by considering the following points:
\begin{itemize}
\item The material is antiferromagnetic (AF) at zero doping. The lower bound of inter-site exchange energy is therefor $-J/4$, where $J$ is the nearest-neighbor exchange constant. 
\item The $d$-wave superconducting state corresponds to a condensate of local singlet pairs~\cite{kotliar1988,zhang1988,yokoyama1988,gros1988}.   The exchange interaction of each singlet pair is strongest if the two electrons or holes occupy nearest neighbor sites. Consequently the lower bound of the exchange energy in the superconducting state is $-3J/4$. 
\item The antiferromagnetic state is suppressed by hole doping, but short-range (in particular nearest-neighbor) spin-correlations persist and diminish gradually as a function of increasing doping.  
\end{itemize}
This scenario provides the formal basis for superconductivity by kinetic energy lowering in the underdoped cuprates~\cite{scalapino1998,hirsch2002,maier2004,toschi2005,haule2007,gull2012,fratino2016}, and for understanding why for the underdoped cuprates the Drude spectral weight increases in the superconducting state~\cite{molegraaf2002,deutscher2005,gedik2005,carbone2006,giannetti2011}. It also explains why this effect flips sign at the overdoped side, and why for overdoped cuprates the superconducting pairing is stabilized by lowering the interaction energy~\cite{levallois2016,gull2012,fratino2016}.

In the absence of interactions with the surrounding sites, the energies of the $\underline{d}d$ excitons are $D_{j}$ ($1 \le j \le 4$). In what follows we assume for simplicity that $D_j$ is independent of doping and temperature.
The excitons can be visualized as a flip of a hole in a $d_{x^2-y^2}$ orbital to one of the other orbitals. Since creating an exciton breaks the spin-correlations with the neighboring sites, the exciton energy is given by $E_j=D_j+X^e-X^g$ where $X^g$ ($X^e$) is the exchange interaction with the neighboring sites of the ground (excited) state. 
For $d_{xy}$, $d_{xz}$ and $d_{yz}$ the symmetry is such that the hopping matrix-element to a nearest-neighbor $d_{x^2-y^2}$ orbital vanishes. Consequently 
the exchange interaction with a hole on a neighboring $d_{x^2-y^2}$ orbital is zero, {\it i.e.} $X^e=0$ so that $E_j=D_j-X^g$. 

For zero doping $X^g\ge -J$ due to the anti-ferromagnetic order and taking into account that there are 4 nearest neighbors, and the lower bound $X^g= -J$ corresponds to full spin polarization. The following expressions for $X^g$ refer to the extremal value for each of the given phases.  At optimal doping $X^g= -3J/4$ due to the d-wave superconducting order. At very high doping $X^g = 0$. Increasing temperature at the underdoped side changes the exchange energy $X^g$ from $-3J/4$ in the superconducting state to $-J$ in the normal state. Correspondingly the exciton energy $E_j=D_j-X^g$ decreases by an amount  $J/4$.  At the overdoped side $X^g$ changes from $-3J/4$ in the superconducting state to $0$ in the normal state, so that the exciton energy {\it in}creases by an amount $3J/4$. Since at high temperature $X^g$ gradually diminishes as a function of doping, having $-J$ and $0$ as the extremal values at zero doping and high doping respectively, a doping level exists where $X^g$ has the same value in the normal and superconducting phase. For this doping there is no superconductivity induced shift of the exciton energy.
In practice, taking into account the experimental limitations on resolution, the superconductivity induced shift of $X^g$ falls below the resolution limit not only at a single doping level but in a range of dopings which, in view of Fig. 5, overlaps with optimal doping.
This state of affairs is depicted in Fig.~\ref{fig:JJ}. 
These effects are strongest for the $\underline{d}_{xy, xz,yz}d_{x^2-y^2}$ excitons. For the  $\underline{d}_{z^2}d_{x^2-y^2}$ exciton all these effects are smaller due to mixing with the oxygen-copper charge transfer excitations. 

We now return to the  question to what extent coupling to excitons plays a role in the pairing mechanism of the cuprates. 
From the above discussion illustrated in Fig.~\ref{fig:JJ} we can conclude that our observations neither exclude nor favor any of the pairing mechanisms described in the introduction.
That said, our experiments expose a strong coupling between charge carriers and $\underline{d}d$ excitons. Therefore, it is  reasonable to expect that electron-exciton coupling contributes to the energy balance of the different phases, including the part of the phase diagram where superconductivity is observed.

\section*{Conclusions}
Using RIXS we have observed clear evidence for the coupling of the low-energy conduction electrons in Bi-2212 to $\underline{d}d$ excitons. The coupling is visible as a distinct change of the $\underline{d}d$ exciton peak at 2~eV when the material becomes superconducting.   
On the basis of cluster calculations the 2~eV peak is predominantly a $\underline{d}_{yz/xz}d_{x^2-y^2}$ exciton. 
While this doesn’t exclude that there exists coupling to the other excitons, their weak intensity makes those couplings harder to detect.
By applying perturbation theory, we can qualitatively understand the experimental temperature dependence at the overdoped side as a renormalization of the $\underline{d}d$ excitons, but this model does not explain the change of sign of this effect as a function of doping. In an approach that takes into consideration that the superconducting pairs and the excitons involve the same electrons, we expect a blueshift of the exciton energy smaller than $3J/4$ when the material turns superconducting.
The change of sign of the superconductivity induced changes from underdoped to overdoped is a natural consequence of the presence (absence) of short-range AF correlations in normal state of the former (latter), which make way for d-wave singlet correlations in the superconducting phase. 

\section*{\bf Acknowledgments} 
We thank A.~Georges, G.~Ghiringelli and X.~X.~Huang for stimulating discussions. The RIXS experiments have been performed at the ADRESS beamline of the Swiss Light Source at the Paul Scherrer Institut (PSI). 
%{\bf Funding:}
We acknowledge funding support from the Swiss National Science Foundation through projects No. 179157 (D.v.d.M.), 178867 (T.S.), and 160765 (T.S.). 
T.P.D. was supported by the U.S. Department of Energy (DOE), Office of Basic Energy Sciences, Division of Materials Sciences and Engineering, under Contract No. DE-AC02-76SF00515.
T.C.A. acknowledges funding from the European Union's Horizon 2020 research and innovation programme under the Marie Sk\l{}odowska-Curie grant agreement No. 701647 (PSI-FELLOW-II-3i program). The work at BNL was supported by the US Department of Energy, office of Basic Energy Sciences, contract No. DOE-sc0012704. 
{\bf Data and materials availability:} The datasets generated and analyzed during the current study are available in Ref.~\cite{opendata}.  These will be preserved for 10 years.

\appendix
\section{Experimental methods}
\label{section:data_acquisition}
Single crystals of Bi$_2$Sr$_2$CaCu$_2$O$_{8-x}$ (Bi-2212) were prepared as described in Ref.~\cite{wen2008} with carrier concentrations $p=0.11$
(underdoped, $T_c=70$~K), $p=0.16$ (optimally doped, $T_c=91$~K), and $p=0.21$ (overdoped, $T_c=70$~K). 
An additional optimally doped sample ($p=0.16$, $T_c=91$~K) was prepared as described in Ref.~\cite{weyeneth2009}. 

\noindent
RIXS spectra were measured at the ADRESS beamline of the Swiss Light Source (Paul Scherrer Institut, Villigen PSI), using $\sigma$ polarization of the incoming x-rays with incoming energy resonant at the Cu $L_3$ edge, $E\simeq931.2$~eV.
The scattering angle between the incoming and outgoing x-rays beams was set at 130 degrees. The sample manipulator allowed adjustment of azimutal and polar angles to a momentum transfer of $\vec{q}=(0.36,0,2)$,  $\vec{q}=(0.25,0.25,2)$
such that the out-of-plane momentum transfer was a multiple of the reciprocal lattice vector $c^*$. The outgoing x-rays were detected without resolving the polarization. The energy resolution was $\Delta E =0.13$~eV (as obtained by measuring the elastic line on a tape containing graphite), which is sufficiently precise for measuring the $\underline{d}d$ exciton at 2~eV while maintaining good statistics.  In order to define the nodal and antinodal directions, the samples were aligned using aa Laue setup. The samples were cleaved at the base temperature ($20$~K) at a pressure of $10^{-10}$~mbar.
  
For each doping and momentum direction, the spectra were acquired in the temperature range of approximately $50$~K below $T_c$ to $50$~K above $T_c$. 
The temperature was swept at a fixed slow rate of $0.2$~K/min ($0.3$~K/min for the optimally doped) while continuously sampling the spectra, resulting in 8 to 10 hours temperature scans.   
We binned the data in intervals of approximately $5$~K in temperature.
To avoid radiation damage on the focus position of the x-ray beam, the beam was slowly scanned along a line of 50$~\mathrm{\mu m}$ during the data acquisition.
We normalized the spectra to the total counts in the energy range from $-1$ to $10$~eV. 
The zero energy position $\omega_0$ was obtained using the following procedure:
For each temperature the spectra are fitted to the expression
\begin{eqnarray}
I(T,\omega)&=& I_0 F(\omega_0,\gamma_0,\sigma_0,\omega) +
\nonumber\\
&+&
\sum_{n=1}^6I_n F(\omega_n+\omega_0,\gamma_n,\sigma_n,\omega)
\end{eqnarray}
where $\omega_0$ describes the zero-energy position of the quasi-elastic peak, $\omega_n$ the relative  energy of all other peaks, and $\gamma_n$ and $\sigma_n$ are the Lorentzian and Gaussian widths of the Voigt profile $F(\Omega,\gamma,\sigma,\omega)$ respectively. An additional weak peak at twice the magnon energy was needed in the fitting procedure to reproduce the asymmetric line-shape in the magnon region.

\begin{figure}[!t]
    \centering
    \includegraphics[width=\columnwidth]{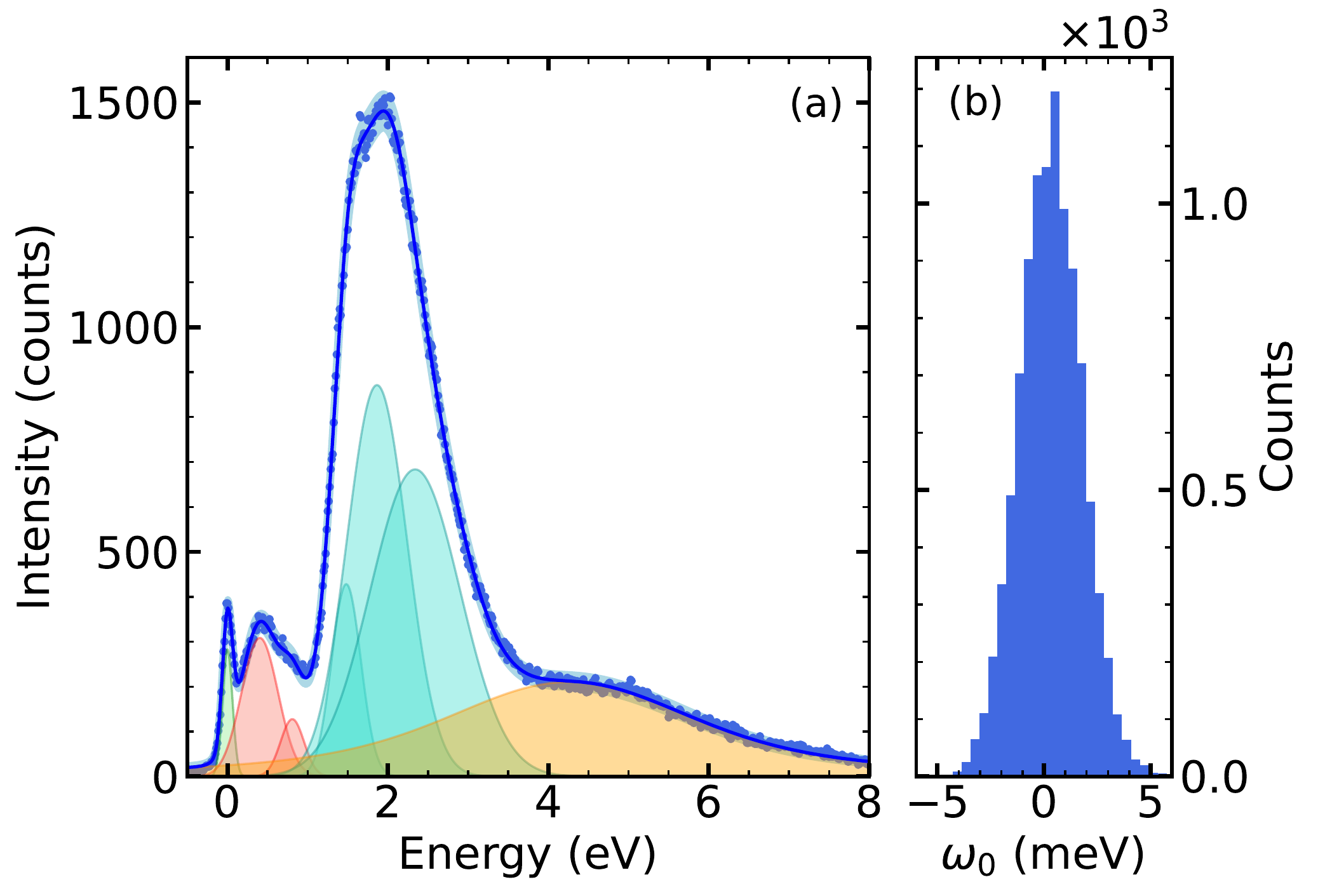}
    \caption{(a) Fitted spectrum for the OD70K $\vec{q}=(0.36,0)$ at 30~K; the shaded area indicates the Poissonian noise. The shaded Voigt peaks represent the decomposition of the spectrum. (b) Distribution of the elastic peak position extracted by fitting $10^4$ spectra with random Poissonian noise.}
    \label{fig:fig_noise}
\end{figure}

\begin{figure}[!t]
    \centering
    \includegraphics[width=0.8\columnwidth]{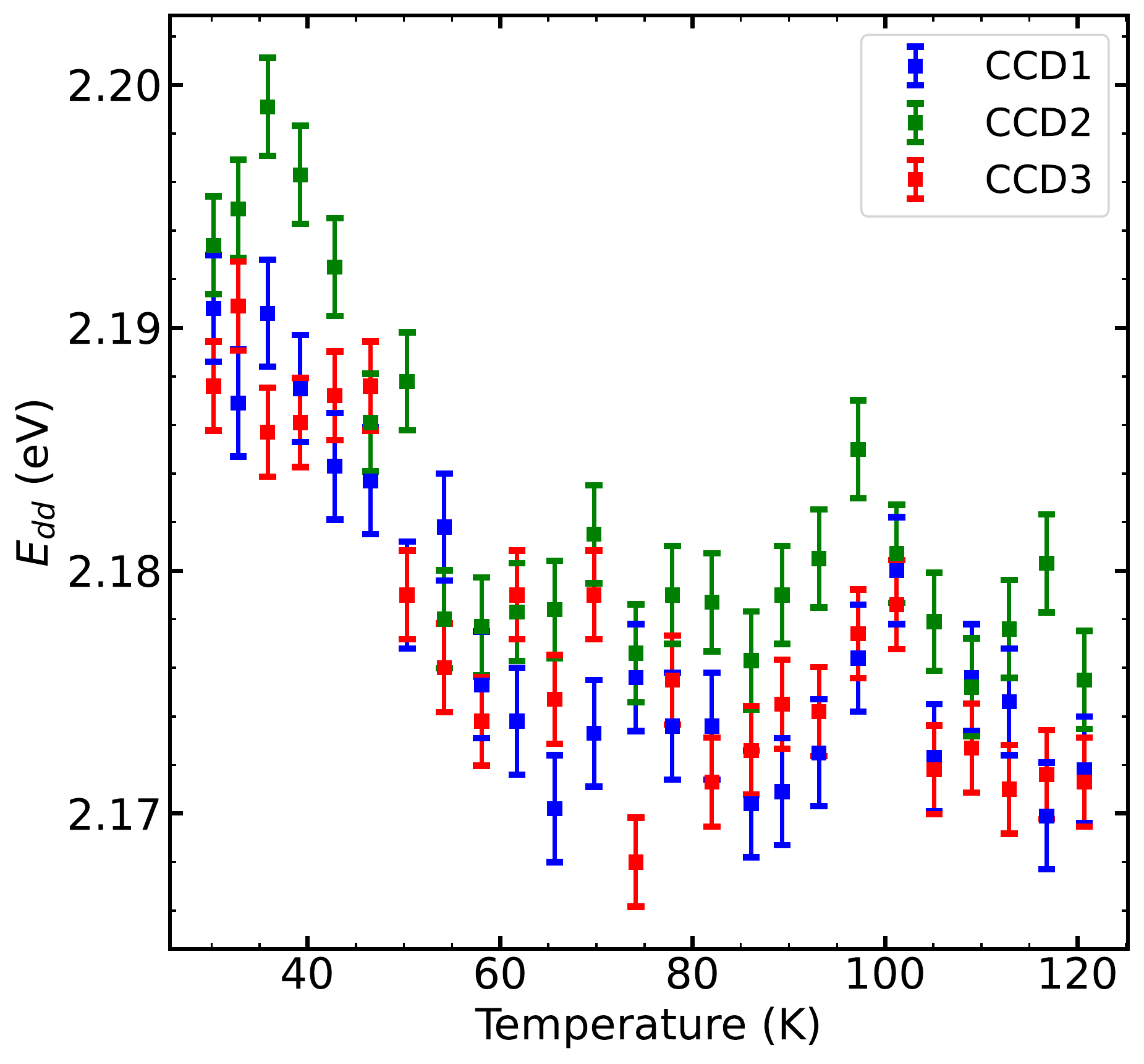}
    \caption{Average energy and of the $\underline{d}d$ exciton peak of overdoped   Bi-2212 ($p=0.21$,$T_c=70$~K) for momentum $(0.36,0)$ measured with 3 different detectors.}
    \label{fig:CCD}
\end{figure}

We used the following method to determine the error bar of the zero energy position $\omega_0$. We have simulated the error in fitting $\omega_0$ due to the statistical noise. To do so, we consider an ideal spectrum $I_\text{fit}(\omega)$ given by the fit at low temperature, such that it reproduces the experimental spectrum for a particular doping and momentum transfer. We show the case of the OD 70K, which is representative for other dopings and momenta, all of which have similar features.
To simulate the statistical noise we add a random Poissonian noise $\delta I_\text{Pn}(\omega)$ to the spectrum of equal magnitude as in the experimental data (\textit{i.e.}\ we simulated the noise level of $N$ counts per channel as $\delta N \sim \sqrt{N}$). We then fitted the noisy spectrum $I_\text{fit}(\omega)+\delta I_\text{Pn}(\omega)$ with the previous starting parameters and repeated the procedure multiple times to generate a statistically significant sample. We considered 10000 realizations, the result of which corresponds to the distribution shown in Fig~\ref{fig:fig_noise}. For this particular example the distribution has a full width at half maximum $3.4$~meV corresponding to $2\sigma=2.9$~meV where $\sigma$ is the standard deviation. This procedure was repeated for each momentum and doping giving the error bars of length $2\sigma$ in Figs.~\ref{fig:first_moment}, and \ref{fig:CCD}.

With incident $\sigma$ polarization, due to direct re-emission the spectra present a reasonably intense elastic peak, which is convenient since all energies are measured relative to the position of the zero-loss peak. A possible complication arises if, given the finite resolution (in the present case around 130~meV) low energy inelastic features such as phonons also contribute to what we should now interpret as a quasi-elastic peak. If these contributions depend on temperature, this imports a temperature dependence in the quasi-elastic peak, which carries over to a spurious temperature dependence of the energy loss of \textit{e.g.}\ the magnon peak and the $\underline{d}d$ excitons. Suzuki \textit{et al.}~\cite{suzuki2018} have reported temperature dependence of the low energy continuum of Bi-2212 at small momentum transfer. On the other hand such contributions were not present in their data for high momentum transfer similar to the momentum values subject of the present study. 
In fact, if present, a spurious temperature dependence of the quasi-elastic peak will affect both the magnon and the $\underline{d}d$ exciton peak in exactly the same way. Here we observe that the temperature dependencies of magnon and exciton energies are completely different (see Fig.~\ref{fig:first_moment}) and the temperature dependence of the magnon is extremely weak. Together, it indicates the absence of a spurious temperature dependence of the quasi-elastic peak for the high momentum transfers considered here, in accordance with the findings of Suzuki \textit{et al.}~\cite{suzuki2018}.

The RIXS spectra are collected by 3 separate CCD detectors. 
The spectra collected with each detector were analyzed separately. 
The zero-energy position of each spectrum was determined as detailed before. 
In Fig.~\ref{fig:CCD} we report the average energy $E$ 
of the $\underline{d}d$ exciton region ($\omega_1=1$~eV and $\omega_2=4$~eV) for the overdoped sample and momentum transfer $(0.36,0)$, measured with the three detectors. Above $\sim 70$~K the average energy is almost temperature independent, and below $\sim  70$~K it gradually shifts to $20$~meV higher energy. 
The observation with three detectors of the same temperature dependence of energy indicates that these features are intrinsic properties of this material. For other momentum transfers and dopings we also observe temperature dependencies of the $\underline{d}d$ energy that are qualitatively similar, but the effects are weaker. 
The focus in this paper is therefore on the temperature shift of the average energy of the $\underline{d}d$ manifold.

\begin{figure}[!h]
    \centering
    \includegraphics[width=0.9\columnwidth]{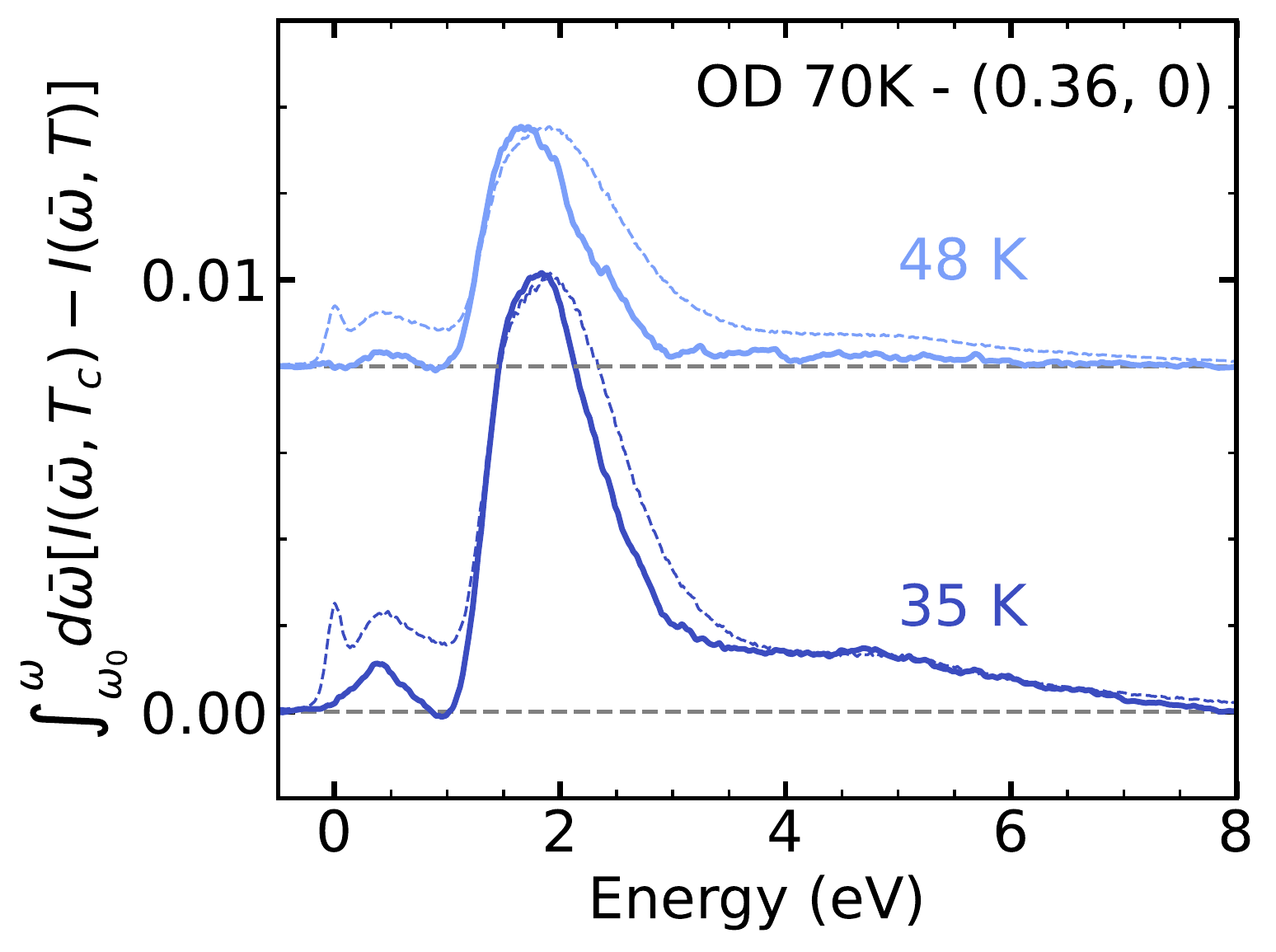}
    \caption{Integral function $G(\omega)$ defined in Eq.~(\ref{eq:G_int}) (solid) and the spectrum $I(\omega, T)$ (dashed) normalized to their maximum, for the OD 70~K sample and $\vec{q}=(0.36, 0)$.}
    \label{fig:IntDiff}
\end{figure}
In Fig.~\ref{fig:IntDiff} we show the integral function defined as
\begin{equation}
    G(\omega)=\int_{\omega_0}^{\omega}\left[I(\omega^{\prime},T_c)- I(\omega^{\prime},T)\right]d\omega^{\prime}
    \label{eq:G_int}
\end{equation}
for the overdoped sample at two temperatures $T$, using the difference spectra reported in Fig.~\ref{fig:spectra} and considering $\omega_0 =-0.5$~eV. The shape matches up to about 2 eV the RIXS spectra at the same temperature $T$. Above 2 eV $G(\omega)$ falls well below $I(\omega)$. Taken together this indicates that the temperature evolution is a shift of  $\underline{d}_{xy}d_{x^2-y^2}$ and $\underline{d}_{xz,yz}d_{x^2-y^2}$ while the shift of $\underline{d}_{z^2}d_{x^2-y^2}$ is significantly smaller.

\section{\boldmath Cluster calculations of the $\underline{d}d$ exciton multiplets}
\label{app:multipletRIXS}
The expression for the RIXS intensity is given by the Kramers-Heisenberg formula
\begin{eqnarray}
&R(\vec{e}_i,\vec{e}_f,\vec{k}_i,\vec{k}_f,\omega_i,\omega_f)=
\frac{1}{Z} \sum_{i,f} e^{-\beta E_i} \nonumber&\\
&\times\left\vert \sum_\nu \frac{\langle f \mid \hat D_{\vec{ k}_f}^*(\vec{e}_f)\mid\nu\rangle \langle\nu\mid \hat D_{\vec{k}_i}({\vec{e}_i})\mid i\rangle}{\omega_i-(E_\nu-E_i)-i \Gamma}\right\vert^2
\times \delta\bigl(\Omega-(E_f-E_i)\bigr)&,
\label{Eq:RIXS}
\end{eqnarray}
which is the largest resonant diagram at lowest order in the fine structure constant $e^2/\hbar c$. Here, $Z$ is the partition function, $E_{i,f}$, $|i,f\rangle$ are the initial and final state energies and eigenstates, respectively, $E_{\nu}$, $|\nu\rangle$ are the RIXS intermediate core hole states, $\Gamma$ is the core hole decay rate, $\vec{e}_{i,f}$, $\vec{k}_{i,f}$, $\omega_{i,f}$ are the incident, scattered x-ray polarization vectors, momenta, and energies, respectively, and $\Omega=\omega_i-\omega_f$. The dipole operator $\hat D$ is evaluated in the Coulomb or ``transverse'' gauge $\nabla\cdot\vec{A}=0$, where $\vec{A}$ is the light vector potential, and the dipole approximation is used.

RIXS is obtained via full matrix diagonalization of the generalized Hamiltonian
\begin{equation}
\hat H_{\mathrm{full}}=\hat H_{\mathrm{int}}+\hat H_{\mathrm{CEF}} + \hat H_{\mathrm{SO}}+\hat H_{\mathrm{KE}}
\label{Eq:full}
\end{equation}
where the terms in the Hamiltonian are given in terms of second quantized operators $c^{\phantom{\dagger}}_{\mu,i,\sigma}, c^\dagger_{\mu,i,\sigma}$ that remove, create electrons in orbital $\mu$ in unit cell $i$ with spin $\sigma$ as follows:

The Coulomb interaction
\begin{equation}
\hat H_{\mathrm{int}} = \frac{1}{2} \sum_i\sum_{\alpha,\beta,\gamma,\delta}\sum_{\sigma,\sigma'} U_{\alpha,\beta,\gamma,\delta} c^{\dagger}_{i,\alpha,\sigma} c^{\dagger}_{i,\beta,\sigma'}
c^{\phantom{\dagger}}_{i,\delta,\sigma'} c^{\phantom{\dagger}}_{i,\gamma,\sigma}
\label{Eq:Hint}
\end{equation}
represent generalized local two-body interactions that can include both core and valence electrons, while the sum of single particle terms can be grouped together as
\begin{equation}
\hat H_{\mathrm{CEF}} + \hat H_{\mathrm{SO}}+\hat H_{\mathrm{KE}} =
\sum_{i,j}\sum_{\sigma,\sigma'}\sum_{\mu,\nu}
t_{i,j,\sigma,\sigma'}^{\mu,\nu} c^\dagger_{i,\mu,\sigma}c^{\phantom{\dagger}}_{j,\nu,\sigma'}
\end{equation}
with the tensor $t$ having site energies, crystal fields and spin-orbit contributions (for $i=j$ terms), and hybridization among orbitals ($i\ne j$).

We utilize a unit cell consisting of five copper $3d$ orbitals, three copper $2p$ core orbitals, and three oxygen $2p$ ligand orbitals along the $x$ and $y$ bond directions, giving 14 orbitals per unit cell. We consider a 2-site cluster (28 orbitals) to yield momentum transfer $(\frac{1}{2},0)$ in reciprocal lattice units. The Hilbert space is restricted to contain at most one core hole in the cluster.

The following set of parameters have been used (hole language is employed). All energies are in units of eV. Cu $3d$ site energies: $(0, 1.8, 0.9, 1.15)$ for $d_{x^2-y^2}$, $d_{z^2}$, $d_{xy}$, $d_{xz,yz}$ orbitals, respectively. O $2p$ site energies: 2.8 for $p_{xx}$, $p_{yy}$, $p_{xy}$, $p_{yx}$ and 0.8 for $p_{xz}$, $p_{yz}$ orbitals, where $p_{i,j}$ denotes ligand orbital $2p_j$ for site $i$ in the unit cell. Slater parameters for Cu $3d$: $F_0=6$, $F_2=0.13$, $F_4=0.025$ to approximately give Hubbard $U=7.4$ and Hund's $J=0.9$. Slater parameters for core Cu $2p$--$3d$ interaction: $F_0=6$, $F_2=0.15$, $G^1=4.63$, $G^3=2.63$. Slater parameters for O $2p$: $F_0=1.0$, $F_2=0.1$. Core spin-orbit coupling: 10.5. Hybridizations: $3d_{x^2-y^2}$--$2p_{x,y}=1$, $3d_{z^2}$--$2p_{x,y}=0.1$, $3d_{xy}$--$2p_{y,x}=0.225$, $3d_{xz,yz}$--$2p_{z}=0.1$. The core hole decay rate $\Gamma=0.1$.

Finally, the incident photon polarization direction is taken to be along $y$ direction ($\sigma$ polarization in the scattering plane), and the outgoing photon is unpolarized. In Fig.~\ref{fig:RIXS1} of the main text we show the peaks resulting from the calculations and the relative intensity normalized to the $yz$ exciton.\\

\section{\boldmath Weak-coupling model for the temperature dependence of the $\underline{d}d$ excitons}\label{section:BCS}
For a two-dimensional single-band superconductor, the temperature-dependent BCS charge susceptibility at wave-vector $\vec{q}$ and complex energy $z$ is
\begin{widetext}
\begin{equation}
	\chi(\vec{q},z)=2\int\frac{d^2k}{(2\pi)^2}
	\int_{-\infty}^{\infty}d\varepsilon_1\int_{-\infty}^{\infty}d\varepsilon_2 \left[A\left(\vec{k},\varepsilon_1\right)A\left(\vec{k}+\vec{q},\varepsilon_2\right)
    +B\left(\vec{k},\varepsilon_1\right)B\left(\vec{k}+\vec{q},\varepsilon_2\right)\right]
    \frac{f\left(\varepsilon_1\right)-f\left(\varepsilon_2\right)}{z+\varepsilon_1-\varepsilon_2}.
\end{equation}
\end{widetext}
The wave-vectors $\vec{k}$ span the first Brillouin zone, $f(\varepsilon)=1/\left[\exp\left(\varepsilon/k_{\mathrm{B}}T\right)+1\right]$ is the Fermi distribution at temperature $T$, and $A$, $B$ are the normal and anomalous spectral functions, respectively. The spectral functions depend on the band dispersion measured from the temperature-dependent chemical potential, $\xi_{\vec{k}}=\varepsilon_{\vec{k}}-\mu$, and on the temperature-dependent superconducting gap function $\Delta_{\vec{k}}$. They are best expressed in terms of the Bogoliubov excitation energies $E_{\vec{k}}=(\xi_{\vec{k}}^2+\Delta_{\vec{k}}^2)^{1/2}$ and coherence factors $u_{\vec{k}}^2=\frac{1}{2}(1+\xi_{\vec{k}}/E_{\vec{k}})$ and $v_{\vec{k}}^2=1-u_{\vec{k}}^2$, according to $A(\vec{k},\varepsilon)=u_{\vec{k}}^2\delta\left(\varepsilon-E_{\vec{k}}\right)+v_{\vec{k}}^2\delta\left(\varepsilon+E_{\vec{k}}\right)$ and $B(\vec{k},\varepsilon)=\frac{\Delta_{\vec{k}}}{2E_{\vec{k}}}\left[\delta\left(\varepsilon-E_{\vec{k}}\right)-\delta\left(\varepsilon+E_{\vec{k}}\right)\right]$.

For the band dispersion, we use a square-lattice tight-binding model with hopping amplitudes up to the fifth neighbors,
\begin{multline}
	\varepsilon_{\vec{k}}=
	2t_1\left[\cos\left(k_xa\right)+\cos\left(k_ya\right)\right]
	+4t_2\cos\left(k_xa\right)\cos\left(k_ya\right)\\
	+2t_3\left[\cos\left(2k_xa\right)+\cos\left(2k_ya\right)\right]\\
    +4t_4\left[\cos\left(2k_xa\right)\cos\left(k_ya\right)+\cos\left(k_xa\right)\cos\left(2k_ya\right)\right]\\
    +4t_5\cos\left(2k_xa\right)\cos\left(2k_ya\right),
\end{multline}
where $a$ is the lattice spacing. The parameters are $t_{1-5}=(-148.8,40.9,-13,-14,12.8)$ meV, as determined in \cite{norman1995}. At each temperature, we adjust the chemical potential $\mu$ such that the hole density is $1-n\equiv\rho$, where the electron density $n$ is
\begin{equation}
	n=2\int\frac{d^2k}{(2\pi)^2}\int_{-\infty}^{\infty}d\varepsilon\,A(\vec{k},\varepsilon)f(\varepsilon).
\end{equation}
The gap function has $d_{x^2-y^2}$ symmetry with an empirical temperature dependence that follows closely the numerical result of the BCS gap equation for a two-dimensional $d$-wave superconductor:
\begin{equation}
    \Delta_{\vec{k}}(T)=\Delta_{\vec{k}}(0)
    \sqrt{1-x^2}\left(1+\frac{x^2}{2}e^{-\frac{4}{5}x^2}\right)
\end{equation}
where $ x={T}/{T_c}$ and the $d$-wave gap reads $\Delta_{\vec{k}}(0)={\Delta_0}\left[\cos\left(k_xa\right)-\cos\left(k_ya\right)\right]/2$.
We set $\rho=0.2$, $T_c=70$~K, and $\Delta_0=30$~meV to represent the overdoped sample.
\begin{figure}[!t]
\centering
\includegraphics[width=0.9\columnwidth]{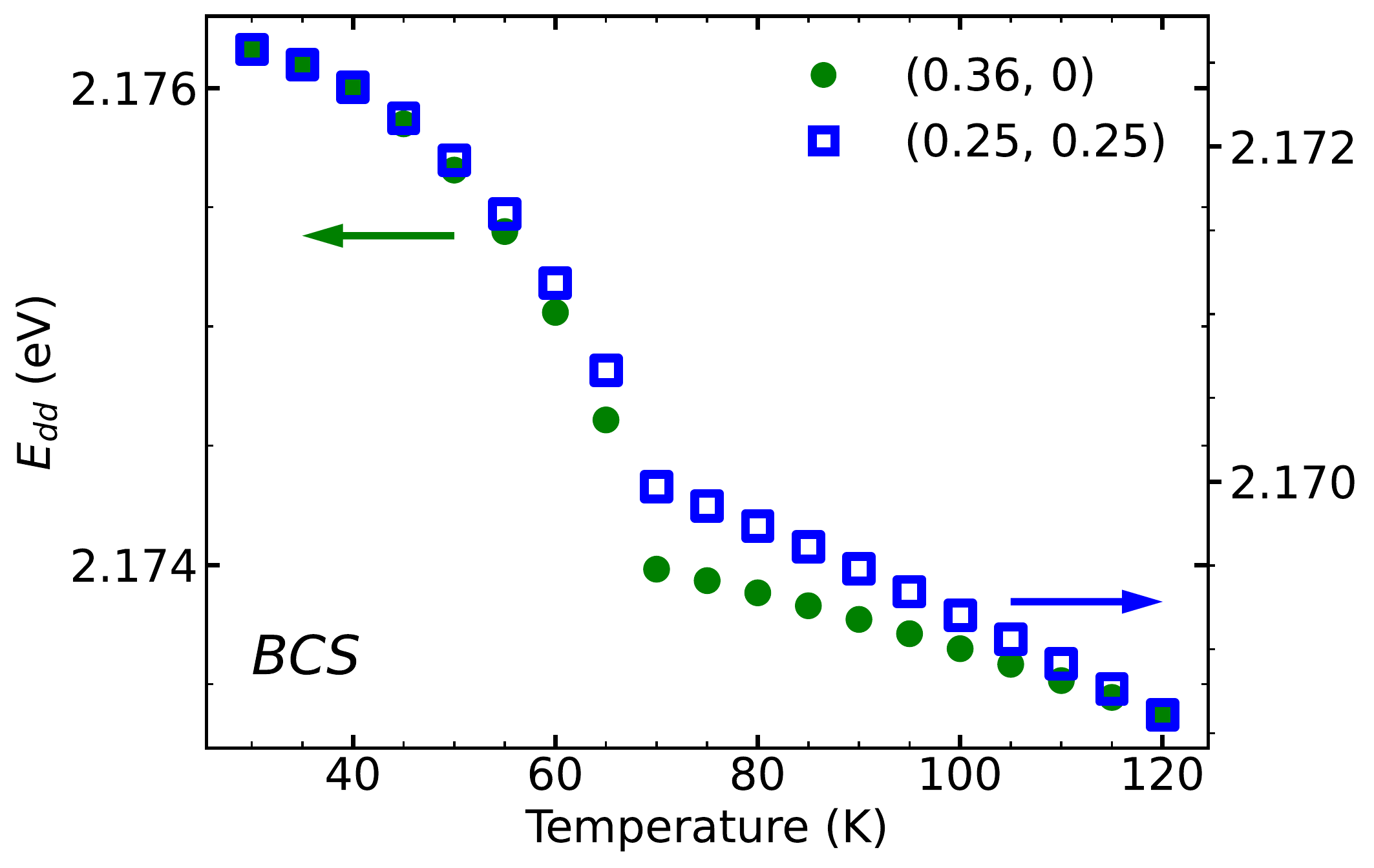}
\caption{Temperature dependence of the $\underline{d}d$ exciton average energy, from the spectral function computed in Eq.~(\ref{eq:Idd}) for the two direction in momentum space $\vec{q}=(0.36, 0)$ (green circles) and $\vec{q}=(0.25, 0.25)$ (blue squares).}
\label{fig:chi}
\end{figure}

For the exciton spectral function the susceptibility enters via the self-energy correction
\begin{equation}\label{eq:Idd}
I_{\underline{d}d}(\vec{q},\omega)\propto -\mathrm{Im}\,\frac{1}{\omega-\omega_{\underline{d}d} -g_{\vec{q}}^2\chi(\vec{q},\omega)+i\Gamma},
\end{equation}
where $\omega_{\underline{d}d}$  is the unrenormalized energy of the exciton and $\Gamma$ represents the exciton intrinsic linewidth--mostly due to interaction with the phonons--that we represent phenomenologically with a constant scattering rate $\Gamma=0.15$~eV.   

We substitute the BCS susceptibility in Eq.~(\ref{eq:Idd}) with $g_{\vec{q}}=2$~eV and we sett $\omega_{\underline{d}d}=1.973$~eV and $\omega_{\underline{d}d}=1.87$~eV for $\vec{q}=(0.36,0)$ and $\vec{q}=(0.25,0.25)$, such that the peak position at $T_c$ is 2.174~eV and 2.17~eV, respectively (see Fig.~\ref{fig:first_moment}). Fig.~\ref{fig:chi} shows the resulting temperature dependence of the exciton energy.
%\clearpage
\providecommand{\noopsort}[1]{}\providecommand{\singleletter}[1]{#1}%
%
%
%\bibliography{references}
%
\end{document}